\documentclass[physrev,twocolumn,superscriptaddress,floatfix]{revtex4-2}

\usepackage{dcolumn}
\usepackage{amsmath}
\usepackage{amssymb}
\usepackage{amsthm}

\usepackage{graphicx}
\usepackage{bm}
\usepackage[T1]{fontenc}
\usepackage{xcolor}
\usepackage{url}
\usepackage[bf]{subfigure}
\usepackage{rotating}
\usepackage{scalefnt}
\usepackage{multirow}
\usepackage{scalefnt}
\usepackage{color}

\newcommand{\E}[1]{\left\langle #1 \right\rangle}

\newcommand{\A}{\mathbf{A}}

\begin{document}

\title{From subcritical behavior to elusive transitions in rumor models}

\author{Guilherme Ferraz de Arruda}
\email{gui.f.arruda@gmail.com}
\affiliation{ISI Foundation, Via Chisola 5, 10126 Torino, Italy}

\author{Lucas G. S. Jeub}
\affiliation{ISI Foundation, Via Chisola 5, 10126 Torino, Italy}

\author{Ang\'{e}lica S. Mata }
\affiliation{Departamento de F\'{\i}sica, Universidade Federal de Lavras, 37200-900, Lavras, Minas Gerais, Brazil}

\author{Francisco A. Rodrigues}
\affiliation{Departamento de Matem\'{a}tica Aplicada e Estat\'{i}stica, Instituto de Ci\^{e}ncias Matem\'{a}ticas e de Computa\c{c}\~{a}o, Universidade de S\~{a}o Paulo - Campus de S\~{a}o Carlos, Caixa Postal 668, 13560-970 S\~{a}o Carlos, SP, Brazil.}

\author{Yamir Moreno}
\affiliation{ISI Foundation, Via Chisola 5, 10126 Torino, Italy}
\affiliation{Institute for Biocomputation and Physics of Complex Systems (BIFI), University of Zaragoza, Zaragoza 50009, Spain}
\affiliation{Department of Theoretical Physics, University of Zaragoza, Zaragoza 50009, Spain}

\begin{abstract}
       Rumor and information spreading are natural processes that emerge from human-to-human interaction. Mathematically, this was explored in the popular Maki--Thompson model, where a phase transition was thought to be absent. Here, we show that a second-order phase transition is present in this model which is not captured by first-order mean-field approximations. Moreover, we propose and explore a modified version of the Maki--Thompson model that includes a forgetting mechanism. This modification changes the Markov chain's nature from infinitely many absorbing states in the classical setup to a single absorbing state. In practice, this allows us to use a plethora of analytic and numeric methods that permit the models' characterization. In particular, we find a counter-intuitive behavior  in the subcritical regime of these models, where the lifespan of a rumor increases as the spreading rate drops, following a power-law relationship. This means that, even below the critical threshold, rumors can survive for a long time.  Together, our findings suggest that the dynamic behavior of rumor models can be much richer than previously thought. Thus, we hope that our results motivate further research both analytically and numerically.

\end{abstract}

\maketitle

Rumor spreading is a spontaneous process that emerges from social interactions and can occur in real or virtual environments. Electronic media has increasingly contributed to the impact of this type of process on people's daily lives. More concretely, these processes include, but are not limited to, propagation of information or gossips and can be even used to model fake news~\cite{Richardson2002, GALAM2003, Kimmel2004, Guerin2006, Davis2020}. Due to its practical relevance, it is essential to understand the evolution mechanism of rumor spreading on heterogeneous networks. Although these examples are easily relatable with our daily lives, surprisingly, this class of processes is much less explored than other stochastic dynamics~\cite{Vespignani2015, Arruda2018}. Indeed, the first rumor model was introduced in 1964 by Daley and Kendall (DK)~\cite{daleykendall1964} and was generalized to networks only in 2004~\cite{moreno2004}. This model assumes that the population is homogeneously connected and the individuals are classified in one of three states: (i) ignorant, who is someone that does not hold the information, (ii) spreader, which is an individual that knows the rumor and is willing to spread it, and (iii) stifler, i.e., someone who knows the rumor but does not spread it. In this model, the transitions are based on contact between individuals. The spreading happens when a spreader contacts an ignorant, and the stifling occurs through the contact between two individuals that are aware of the rumor. In 1973, Maki and Thompson (MT)~\cite{makithompson1973}, considered the same spreading scheme and homogeneous assumptions, slightly changing the annihilation mechanism by considering directed contacts. In other words, when a spreader contacts another spreader or a stifler, just the individual that initiated the contact turns into a stifler.

Mean-field approaches have proved to be useful for understanding the behavior of spreading processes in complex networks~\cite{Boguna2002, moreno2004, Nekovee2007, Barrat08:book, Mieghem2009, VanMieghem12, Mieghem2014, Vespignani2015, Arruda2017b, Arruda2018}. In Ref.~\cite{moreno2004}, Moreno and collaborators used such an approach and showed that the MT rumor model in homogeneous structures does not have a phase transition. This behavior is very intriguing, especially if compared with the SIR (Susceptible-Infected-Recovered) epidemic spreading processes. At first glance, they are similar as both have infinitely many absorbing states. However, the two models have very different annihilation mechanisms. In rumor models this mechanism is driven by the contact between individuals aware of the information, while in the epidemic case it is spontaneous. The epidemic model is very well characterized, with dynamics that converge exponentially to the absorbing state in the subcritical regime~\cite{Mieghem2016} and a well-defined phase transition~\cite{Barrat08:book, Mieghem2009, VanMieghem12, Mata2013, Mieghem2014, Vespignani2015, Mieghem2016, COTA2018, Arruda2018}. Surprisingly, the same can not be said about rumor models in networks as, to the best of our knowledge, up to now the results found in Ref.~\cite{moreno2004} were not challenged nor formally proven. Perhaps, due to the similarities with the epidemic processes, this model remained underexplored. However, one must not forget that the mean-field employed in Ref.~\cite{moreno2004} assumes that the states of the nodes are independent, which, as we will show, is not reasonable in rumor models. Here, we revisit these results using Monte Carlo simulations, showing that the MT model presents a phase transition and a very particular power-law temporal behavior in its subcritical regime, where the survival time of a rumor diverges as the inverse of the spreading rate. Moreover, we propose a small modification in the standard model, changing the process from infinitely many absorbing states to a single absorbing state (the entirely ignorant population), which preserves the MT model's essence, but allows the use of additional analytical and simulation tools. 

\begin{figure}[t!]
    \includegraphics[width=\linewidth]{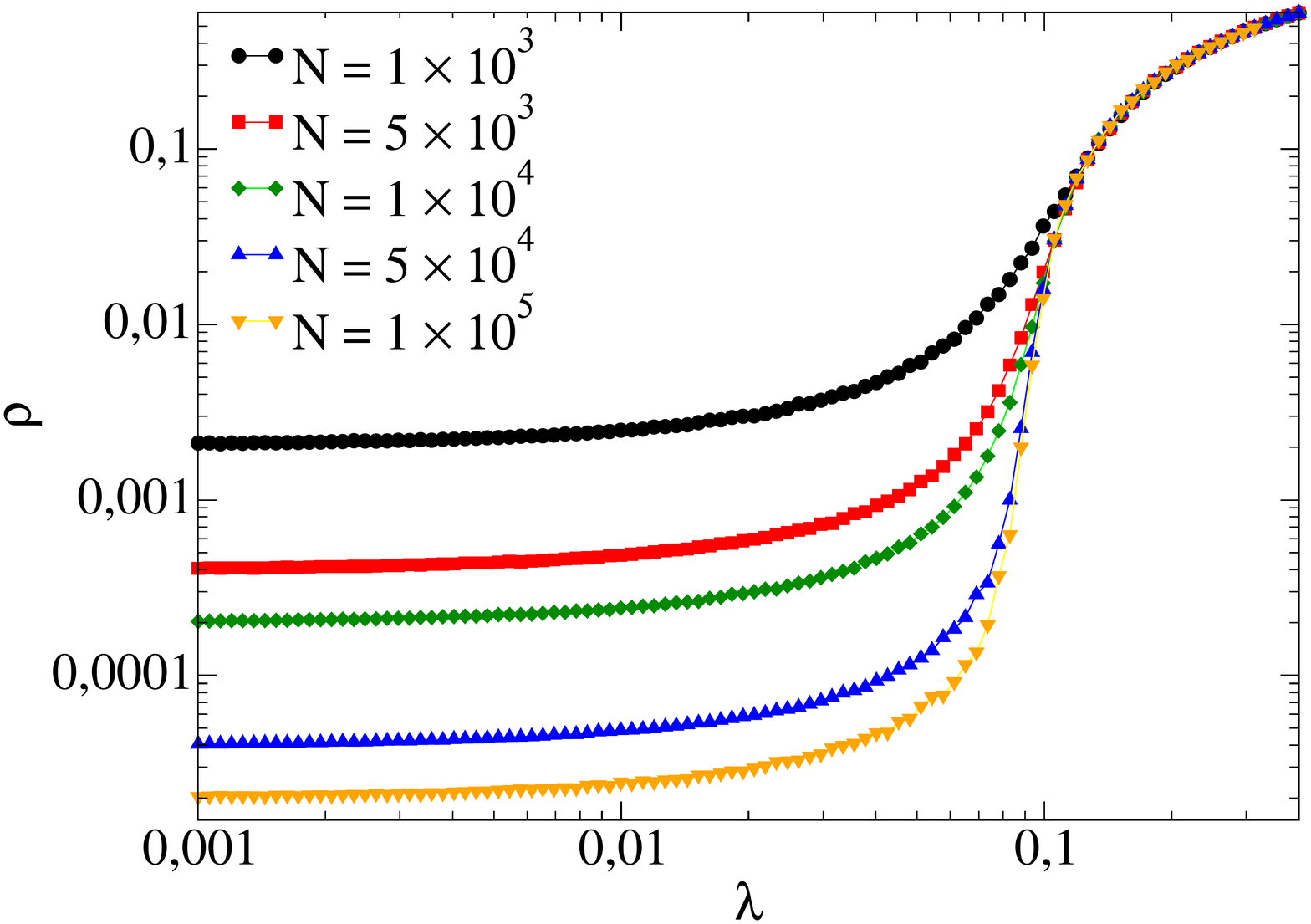}
    \caption{Phase diagram for the standard MT model, $\alpha = 1$ and different sizes on a random regular networks with $\langle k \rangle = 10$.}
    \label{fig:MK_critical}
\end{figure}

Let us first define our model. In alignment with the DK and MT models, here we also have the same set of states (ignorant, spreader, or stifler), which are modeled by associating to each node $i$ three Bernoulli random variables, $X_i$, $Y_i$, and $Z_i$. If node $i$ does not know the rumor,  it is classified as an ignorant ($X_i = 1$). If it knows the rumor and is spreading it, it is called a spreader ($Y_i = 1$). However, if it knows the rumor but does not spread it, it is classified as stifler ($Z_i = 1$). Note that $X_i + Y_i + Z_i = 1$. The spreading evolves through the contact between nodes defined by an undirected network, which is codified by the adjacency matrix $\A$, whose components $\A_{ij}$ are equal to one if there is an edge between nodes $i$ and $j$ and zero otherwise. Our process is defined in continuous-time as a collection of Poisson processes. If the contact is between a spreader and an ignorant, the second node will learn the rumor and become another spreader at rate $\lambda$. On the other hand, if the contact happens between a spreader and someone that already knows the rumor (spreaders or stiflers), then the spreader that initiated the contact will lose interest in the rumor, thus becoming a stifler at a rate $\alpha$. Also, we introduce a forgetting mechanism, where each stifler spontaneously becomes ignorant with a rate of $\delta$. We remark that, if $\delta = 0$, we recover the MT model. Moreover, $\delta$ can be interpreted as a forgetting mechanism or, for evolving rumors, individuals who do not know the rumor's current state. 
These local rules are expressed as
 \begin{equation*} \label{eq:rules}
 \begin{aligned}
  Y_i + X_j & \xrightarrow{\lambda} Y_i + Y_j,\\
  Y_i + Y_j & \xrightarrow{\alpha} Y_i + Z_j,\\
  Y_i + Z_j &\xrightarrow{\alpha} Z_i + Z_j,\\
  Z_i &\xrightarrow{\delta} X_i.\\
 \end{aligned}
 \end{equation*}

\begin{figure}[t!]
    \includegraphics[width=\linewidth]{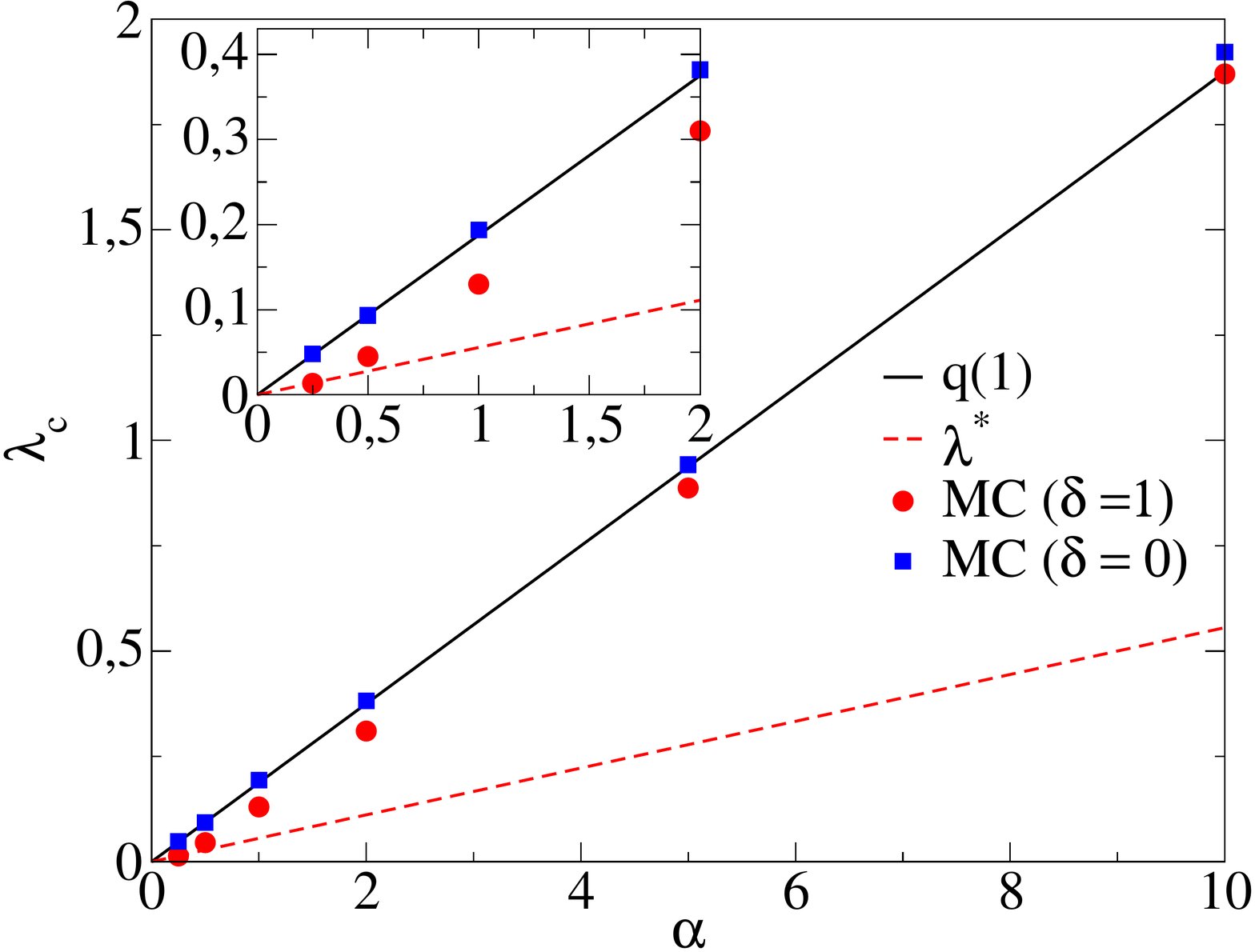}
    \caption{Comparison between analytical and Monte Carlo critical point estimations for random regular networks with $\langle k \rangle = 10$ and $\delta = 1$ and $N = 10^6$.}
    \label{fig:Critical}
\end{figure}

To analyze the model, first, we need to define a phase transition and the critical point in our context. We remark that both the MT and SIR have infinitely many absorbing states. The critical point is the parameter that separates two scaling regimes. Before this point, the number of stiflers (or recovered in the SIR) do not scale with the system size. Hence its fraction goes to zero in the thermodynamic limit. After the threshold, the number of stiflers (recovered) scales with the system size. On the other hand, for processes like SIS (Susceptible-Infected-Susceptible), where we have a single absorbing state (disease-free state), the threshold is defined as the point where above it we have an active state in the thermodynamic limit, and below it, the system goes to the absorbing state. Note that our model has infinitely many absorbing states if $\delta = 0$ and a single absorbing state if $\delta > 0$.

Our first main result is showing that the MT model has a phase transition. Fig.~\ref{fig:MK_critical} shows the average fraction of stiflers $\rho$ as a function of $\lambda$ in the standard MT model ($\delta = 0.0$) with $\alpha = 1.0$ for random regular networks with different sizes. We observe that, as the system size increases, for small $\lambda$ the fraction of stiflers decreases with system size, whereas for large $\lambda$ it is larger than zero and independent of the system size. These results indicate a phase transition in the dynamic behavior of the MT model as a function of $\lambda$.  Although these results are in striking contradiction with the mean-field approximations~\cite{moreno2004, Barrat08:book, Vespignani2015, Arruda2018}, we highlight that the mean-field approach neglects correlations between node states which are crucial in rumor models given the contact-driven stifling mechanism. Also, in Ref~\cite{Arruda2018}, Arruda {\it et al.} quantified the error for this mean-field approximation, showing that a first-order approximation is not accurate for rumor models. 

To better understand this phenomenon, we move to our modified model and concentrate our efforts on the $\alpha \gg \delta$ regime, which includes the MT model. In a transient period, we can neglect the forgetting transitions and assume that there are only two competing mechanisms, the spreading and the stifling. For locally tree-like networks near the absorbing state, we can estimate the expected number of newly created informed nodes $q_k(1)$ that result from the initial spread of the rumor \footnote{At least one spreading event is required to reach the absorbing state from an initial state with one informed node due to the contact-driven stifling process.} to a node of degree $k$ with the following recurrent expression
\begin{equation}
    \scalefont{0.8}
    q_k(i) = \begin{cases}
         \frac{(k-i)\lambda}{i\alpha + (k-i)\lambda} \left[ i \frac{(i+1)\alpha}{(i+1)\alpha + (k-i-1)\lambda} + q_k(i+1) \right] &\text{if } i < k \\
         0 &\text{otherwise}.
         \end{cases}
\end{equation}
By averaging over the degree distribution, the condition that establishes the transition is
\begin{equation}
 q(1) = \langle q_k(1)\rangle_k > 1,
\end{equation}
which is not a closed expression, but can be numerically solved. In Fig.~\ref{fig:Critical} we compare the solution of $q(1)$ with Monte Carlo critical point estimations for random regular networks with $\langle k \rangle = 10$ for both $\delta = 0$ and $\delta = 1$ (see Appendix~\ref{app:MC} for the simulation details). In the same figure, we also present the naive estimation, considering a first-step approximation, given as $\lambda^* = \frac{\alpha}{2 (\langle k \rangle - 1)}$. We observe that $q(1)$ seems to be a reasonable approximation as its precision increases with $\alpha$. On the other hand, $\lambda^*$ might be a reasonable approximation for small enough $\alpha$. These solutions suggest that the process can not be reduced to a first-step approximation, and frustrated trials to get to the absorbing state are expected and should be accounted for. A crucial difference between $q(1)$ and the mean-field approaches that fail to predict a phase transition is that $q(1)$ takes into account that the rate at which a spreader turns into a stifler increases with the number of neighbors to which it spreads the rumor. 

We also analyzed the critical properties of power-law networks, and show the results in Appendix~\ref{app:PL}. Importantly, we observe that in power-law networks, $P(k) \sim k^{-\gamma}$, the threshold seems to vanish for $\gamma < 3.0$, and converges to a non-null value for $\gamma > 3.0$. This behavior is at odds with the behavior of the SIS model, in which the power-law networks have a vanishing critical point for any value of $\gamma$~\cite{chatterjee2009, montford2013, Arruda2018}. Conversely, it is similar to the SIRS (Susceptible-Infected-Recovered-Susceptible) model~\cite{Ferreira2016}, contact process \cite{cp}, the generalized SIS model with weighted infection rates~\cite{KJI} and also modified versions of the SIS model~\cite{COTA2018}. In all of these models, the phase transition can be associated with a collective phenomenon which involves the activation of the whole network, whereas the phase transition in the standard SIS model shows an unusual behavior related to mutual reinfection of hubs~\cite{Boguna2013,Ferreira2016b,Castellano2020,Huang2020}. 

\begin{figure}[t!]
    \includegraphics[width=\linewidth]{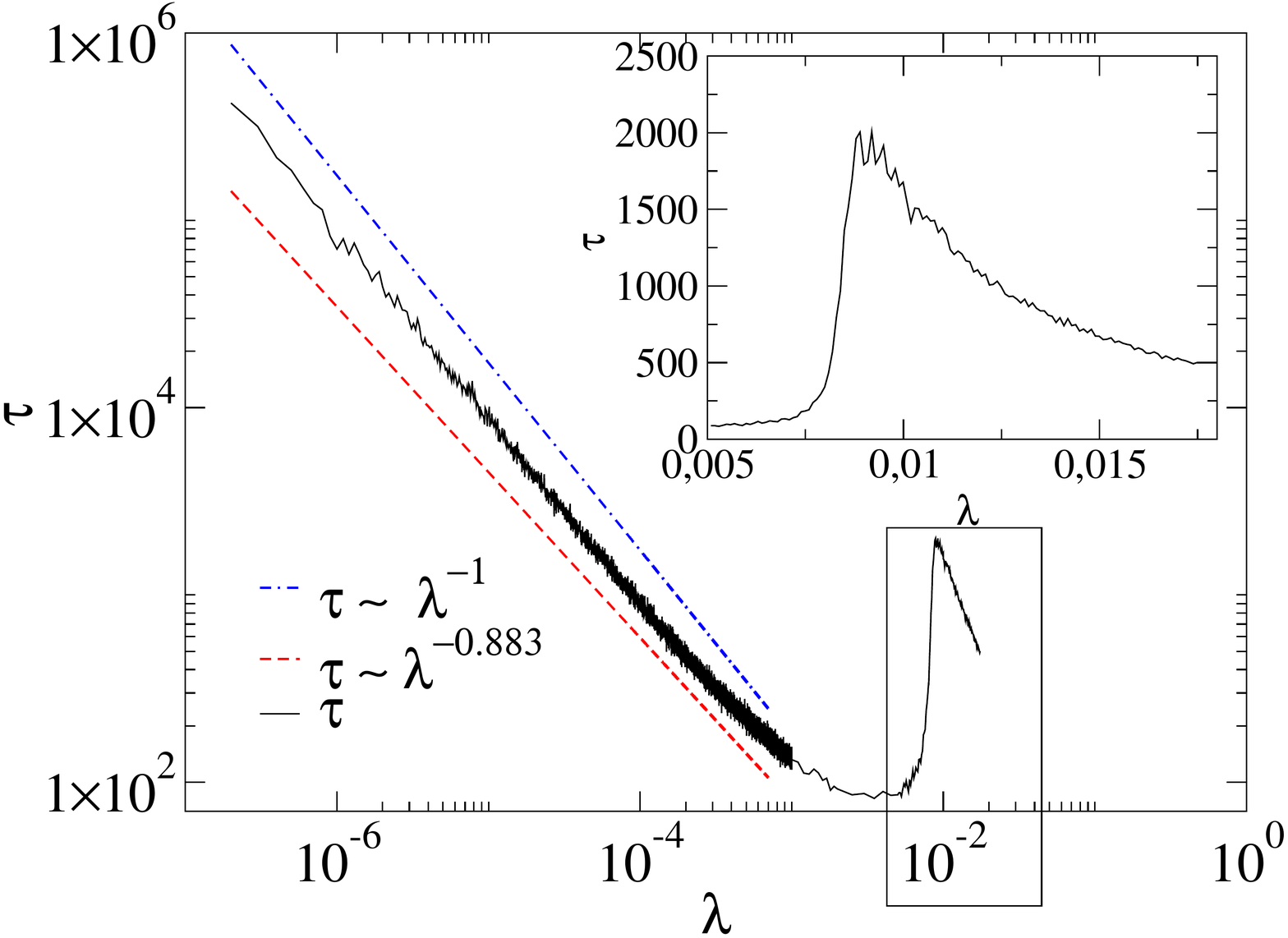}
    \caption{Lifespan as a function of the spreading rate $\lambda$ for $\delta = 1$, $\alpha = 0.5$ on an uncorrelated power-law network with $P(k) \sim k^{-\gamma}$ with $\gamma \approx 2.25$, $N=10^6$. In the main panel we show a wide range of $\lambda$, emphasizing the sub-critical behavior, while in the inset we show the peak that suggests a second-order phase transition. The blue curve (dot dashed line) follows $\tau \sim \lambda^{-1}$ and the red curve (dashed line) follows $\tau \sim \lambda^{-0.8833}$, obtained from a fitting.}
    \label{fig:lifespan}
\end{figure}

\begin{figure}[t!]
    \includegraphics[width=\linewidth]{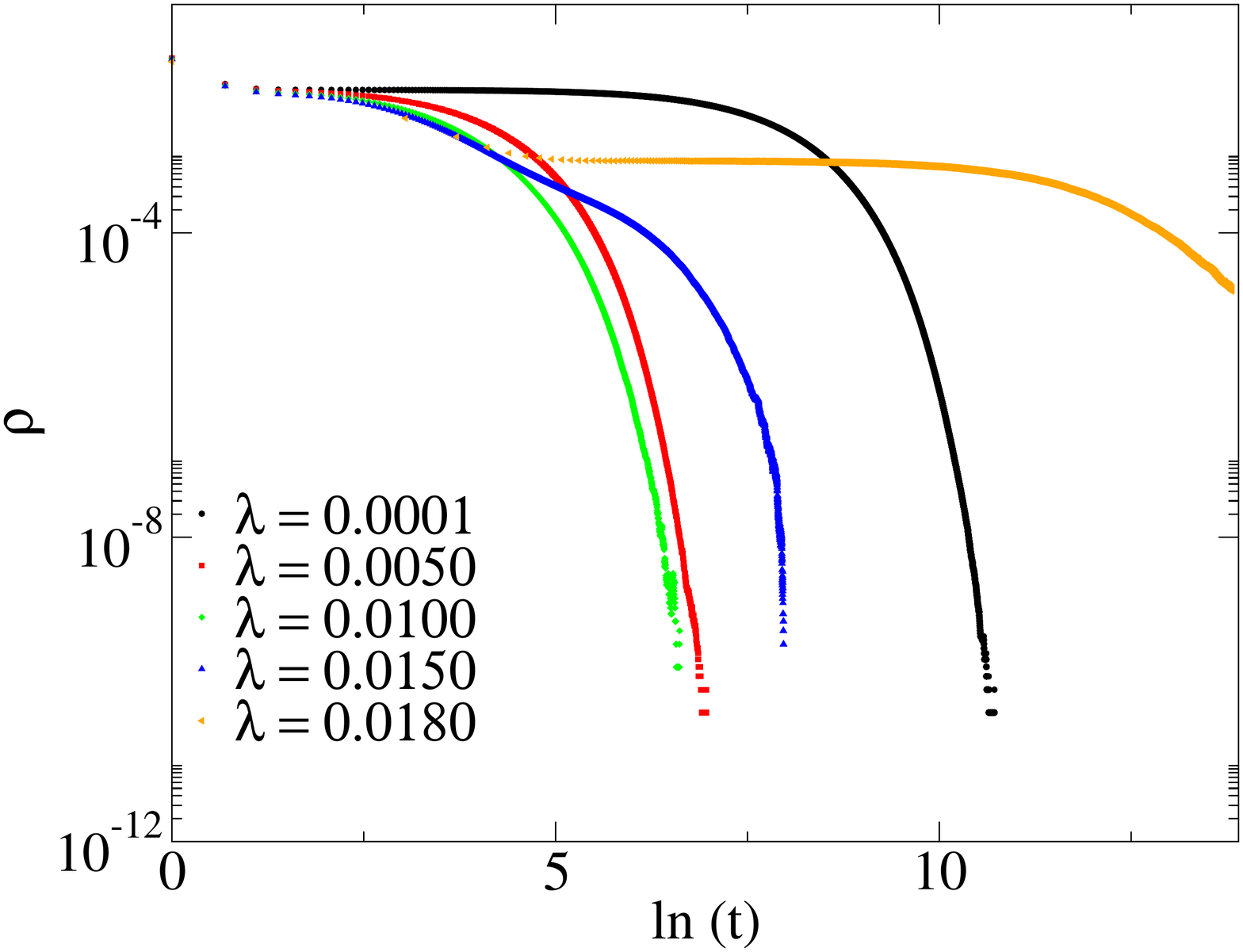}
    \caption{Temporal behavior of the density of spreaders for an uncorrelated power-law network with $N = 10^6$, $\gamma = 2.25$, $\alpha = 0.5$ and $\delta = 1$ in the regime $\lambda \gg \delta$ for values of $\lambda$ near the critical point $\lambda_c \approx 0.015$.}
    \label{fig:subcritical_behavior}
\end{figure}

Another important result regards the subcritical regime, where the rumor can last for a ``very long'' time in the small $\lambda $ regime. This phenomenon was also found in the MT model but, to the best of our knowledge, remained unexplored until now. These behaviors can be observed in Fig.~\ref{fig:lifespan}, which shows the lifespan $\tau$ as a function of the spreading rate $\lambda$ for an uncorrelated power-law network with $P(k) \sim k^{-\gamma}$ with $\gamma \approx 2.25$ and $N=10^6$. In the main figure, we show the subcritical regime, whose lifespan follows $\tau \sim \lambda^{-\eta}$, while in the inset, we show the peak for $\tau$, which suggests a second-order phase transition.  To calculate the lifespan we used the method proposed by Bogu{\~n}{\'a} and collaborators~\cite{Boguna2013}. The simulations starting from a single occupied (spreader) node. For $\lambda$ below the critical point, all realizations usually have a finite and very short lifespan $\tau$. As $\lambda$ grows the average duration of finite realizations increases, diverging at $\lambda_c$. However, for $\lambda > \lambda_c$, the realizations that remain finite have necessarily a short lifespan since the probability of a realization to be endemic increases.  So, $\langle \tau \rangle$ also  decreases as $\lambda$ is increased, in this range of $\lambda$ values. Then, $\langle \tau \rangle$ exhibits a peak that converges to $\lambda_c$ in the thermodynamic limit.

Complementarily, Fig.~\ref{fig:subcritical_behavior} shows the temporal behavior of the density of spreaders $\rho$, where we can see that the time to reach the absorbing state is larger for $\lambda = 0.0001$ than for $\lambda = 0.01$, at odds with exponential decay. Phenomenologically, the transition from the spreader to the stifler depends on at least two individuals who are not ignorant. In the $\lambda \ll \delta$ regime, the forgetting mechanism reduces the number of stiflers faster than the creation of new spreaders, thus increasing the number of frustrated trials to reach the absorbing state. 

\begin{figure}[t!]
    \includegraphics[width=\linewidth]{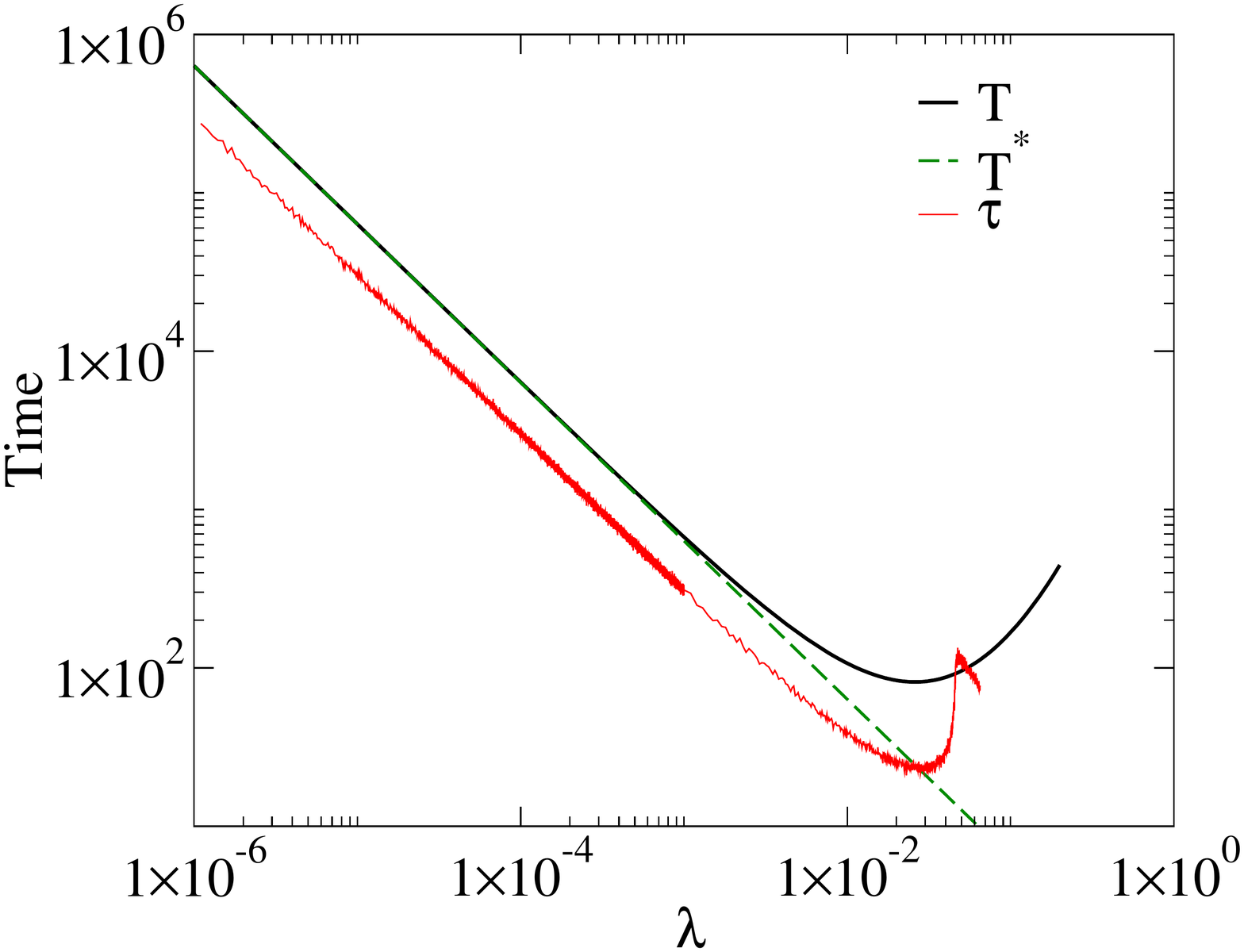}
    \caption{Approximations $T$ and $T^*$ for the time to reach the absorbing state as a function of $\lambda$ for $\delta = 1.0$ and $\alpha = 0.5$. The red curve is the result of Monte Carlo simulations in a random regular network with the same parameters and $N  = 10^5$.}
    \label{fig:Tabs_loglog}
\end{figure}

The subcritical regime can be characterized in terms of the time to reach the absorbing state, $T_{abs}$. This quantity can be approximated under the assumption that there is a single spreader and the population is locally homogeneous, where nodes have $\langle k \rangle$ neighbors. In this case, assuming that node $i$ is a spreader, the shortest sequence of events that will drive the process to the absorbing state consists of: (1) node $i$ spreads the information to node $j$, (2) either $i$ or $j$ turn into stifler, (3) the other node turns into a stifler, (4) either $i$ or $j$ turn into ignorant, and (5) the other node turns into an ignorant. Since our process is Markovian, the respective probabilities $P$ and expected times $\langle \tau \rangle$ of these events are 
\begin{align}
  P_1 &= 1 & \E{\tau_1} &= \frac{1}{\langle k \rangle \lambda}  \nonumber\\
  P_2 &= \frac{2\alpha}{2\alpha + 2\lambda (\langle k \rangle -1)} & \E{\tau_2} &= \frac{1}{2\alpha + 2\lambda (\langle k \rangle -1)}  \nonumber\\
  P_3 &= \frac{\alpha}{\alpha + \lambda (\langle k \rangle -1) + \delta} & \E{\tau_3} &= \frac{1}{\alpha + \lambda (\langle k \rangle -1)} \nonumber\\
  P_4 &= P_5 = 1 & \E{\tau_4} &= \E{\tau_5} = \frac{1}{\delta}. \nonumber
\end{align}
Therefore, the probability and the expected time for reaching the absorbing state through this chain are
\begin{align}
    P_{abs} &= \frac{\alpha^2}{(\alpha+(\langle k \rangle -1) \lambda) (\alpha+\delta+(\langle k \rangle -1) \lambda)}\\
    \langle \tau_{1\rightarrow 5} \rangle &=  \frac{1}{k\lambda} + \frac{2}{\delta} + \frac{3}{2\alpha + 2\lambda (\langle k \rangle -1)}.
\end{align}
From these quantities, we can approximate the average value of $T_{abs}$. In the $\lambda \ll \delta$ and $\lambda \ll \alpha$ regime, $\E{\tau_1} \gg \E{\tau_\ell}$, for $\ell = 2, 3, 4, 5$. Thus, the time spent in the frustrated trials can be approximated by $\E{\tau_1}$. So, $\E{T_{abs}}$ can be approximated by counting the number of times the process fails to reach the absorbing state plus the time it succeed, which is given by
\begin{equation} \label{eq:Tabs}
    T = \sum _{i=1}^{\infty } i \E{\tau_1} (1-P_{abs})^i + P_{abs} \E{\tau_{1\rightarrow 5}} \approx  \E{T_{abs}},
\end{equation}
which solves as
\begin{widetext}
    \scalefont{0.78}
\begin{align} \label{eq:Tabs_exp}
    T = \frac{(\alpha+(\langle k \rangle -1) \lambda) (\alpha+\delta+(\langle k \rangle -1) \lambda) \left((\langle k \rangle -1) \lambda (2 \alpha+\delta)+\alpha \delta+(\langle k \rangle -1)^2 \lambda^2\right)}{\alpha^4 k \lambda}
    +\frac{\alpha^2 \left(\frac{1}{\alpha+\delta+(\langle k \rangle -1) \lambda}+\frac{1}{2 \alpha+2 (\langle k \rangle -1) \lambda}+\frac{2}{\delta}+\frac{1}{k \lambda}\right)}{(\alpha+(\langle k \rangle -1) \lambda) (\alpha+\delta+(\langle k \rangle -1) \lambda)},
\end{align}
\end{widetext}
and for the regime $\lambda \ll 1$ and $\lambda \ll \alpha, \delta$, follows
\begin{equation}
    T^* = \left( \frac{(\alpha + \delta)\delta}{\langle k \rangle \alpha^2} \frac{\alpha}{\langle k \rangle (\alpha + \delta)} \right) \lambda^{-1}.
\end{equation}
Note that Eq.~\eqref{eq:Tabs_exp} is not defined for the standard MT model, i.e., $\delta = 0$. In this case, in the regime $\lambda \ll \alpha$, we expect that $T \sim \lambda^{-1}$ and only two nodes learn the rumor. Fig.~\ref{fig:Tabs_loglog} shows $T$ and $T^*$ as a function of $\lambda$, where the subcritical behavior is dominated by the time spent on the spreading events. Although our assumptions do not cover heterogeneous networks, $T$ has a similar behavior as the one observed in Fig.~\ref{fig:lifespan} for a power-law network, suggesting that our assumptions are reasonable and might be applicable in different substrates. Complementarily, see Appendix~\ref{app:rrn}, where we show $T_{Abs}$ for $\delta = 0$ and the same networks as in Fig.\ref{fig:MK_critical}.

From the analytical viewpoint, the analysis of arbitrary networks is much more evolved, and even the proof of existence of a critical point is impossible in a first-order approximation. However, this argument might be of interest to our community as it points to interesting future directions. In a first-order approximation we assume that the probabilities are independent. Denoting $\E {X_i}$, $\E {Y_i}$ and $\E {Z_i}$ by $x_i$, $y_i$ and $z_i$, respectively, we have
\begin{equation} \label{eq:qmf}
    \scalefont{0.95}
    \begin{cases}
        \frac{d x_i}{dt} =& \delta z_i - \lambda \sum_{k=1}^N \A_{ki} x_i y_k \\
        \frac{d y_i}{dt} =& \lambda \sum_{k=1}^N \A_{ki} x_i y_k - \alpha \sum_{k=1}^N \A_{ki} y_i \left(y_k + z_k \right) \\
        \frac{d z_i}{dt} =& -\delta z_i + \alpha \sum_{k=1}^N \A_{ki} y_i \left(y_k + z_k \right). 
    \end{cases}
\end{equation}
Here we follow an asymptotic analysis, considering that $y_i = y_i^{(1)} \epsilon^c + O(\epsilon^{2c})$, $z_i = z_i^{(1)} \epsilon^k + O(\epsilon^{2k})$ and $x_i \in O(1)$, where $\epsilon \ll 1$. Moreover, we have to consider a scaling of the parameters as $\lambda = \tilde{\lambda} \epsilon^m$ and $\alpha = \tilde{\alpha} \epsilon^n$ to be able to balance the equations. Without loss of generality, we also assume $\delta = 1$. In practice, the scaling of the parameters will be related with the scaling of the nodal probabilities, allowing us to understand the behavior of our model. From Eq.~\eqref{eq:qmf} in the steady-state and neglecting the higher-order terms, we have
\begin{align}
    \scalefont{0.95}
    \begin{cases}
        \delta z_i^{(1)} \epsilon^k &= \tilde \lambda \epsilon^m \sum_{k=1}^N \A_{ki} x_i^{(1)} y_k^{(1)} \epsilon^c\\
        \delta z_i^{(1)} \epsilon^k &= \tilde \alpha \epsilon^n \sum_{k=1}^N \A_{ki} y_i^{(1)} \epsilon^c \left(y_k^{(1)} \epsilon^c + z_k^{(1)} \epsilon^k \right).
    \end{cases}
\end{align}
This relation imposes that $k = n + c + \min(c, k)$ and $k = m + c$, establishing three different possible regimes: (i) $0 < c < k$, where $c = m -n$ and $k = 2m -n$, (ii) $c = k$, where $m = 0$ and $n = -k$, and (iii) $0 < k < c$, where $c = -n$ and $k = m - n$. In the first regime, we have that $y_i^{(1)} = \frac{\tilde{\lambda}}{\tilde{\alpha}}$ and hence $y_i \sim \frac{\lambda}{\alpha}$. For the second and third regimes we have to assume a regular network (i.e., all nodes have the same degree) to obtain a closed-form solution, resulting in $y_i \sim \frac{\lambda}{\alpha \left( \lambda  \Lambda_{\max} + 1 \right)}$ for the second, and $y_i \sim \frac{1}{\alpha \Lambda_{\max}}$ for the third regime. We leave the details of the asymptotic analysis to Appendix~\ref{app:aa}. Note that in the three possible regimes we have positive and non-zero solutions, thus implying that in the first-order approximation the phase transition is not captured. However, this transition was observed in our numerical experiments. For this reason, we call this an elusive transition.

We studied both the standard MT model and a modified version with an additional transition from stifler to ignorant. This simple modification completely changes the thermodynamic behavior of the model. In the MT model, we have infinitely many absorbing states, while in our modified version, we have a single absorbing state~\footnote{The exception is $\delta = 0$, in which we recover the MT model.}. First, we show that the MT model has a phase transition in its dynamic behavior as a function of the spreading rate. In other words, for a one seed initial condition and controlling only the spreading parameter $\lambda$, there are two scaling regimes, one where the fraction of stiflers goes to zero in the thermodynamic limit and a regime where it scales with the population size. This contradicts the commonly accepted result that such a transition is absent in the MT model, which was based on the behavior of first-order mean-field approximations. In our modified model with $\delta > 0$, we have a single absorbing state, and the transition can be found between the rumor-free state and an active state. We provided an expression for locally-tree-like homogeneous networks that allows us to estimate the critical point, covering both the MT and modified models. Crucially, this expression explicitly accounts for the local correlations between states that are ignored by the first-order mean-field approximations that fail to capture the transition. We characterized this transition both in power-law and random regular networks, showing that for $\gamma < 3.0$ the critical point seems to vanish in the thermodynamic limit, while for $\gamma \geq 3$ it converges to a non-null value. These findings seem to be robust against variations in $\alpha$. However, for fixed network size, as we increase $\alpha$, the critical point also increases. More interestingly, we observed a particular subcritical regime, whose lifespan follows $\tau^{-\eta}$ in the $\lambda \ll \delta$ regime, which we studied both analytically and numerically. Phenomenologically, the annihilation mechanism depends on the contacts and, if $\lambda$ is very small, the rumor needs a very long time to reach an absorbing state. Note that, if $\delta > 0$, the rumor might still wander around in the network before die-out. From a practical viewpoint, this means that, in order to contain the rumor spreading, people need to be aware of it. We hope that our findings will motivate further research to better characterize the subcritical regime and phase transition in arbitrary networks analytically, and experiments that might validate the peculiar dynamics of rumor spreading implied by our findings in real-world systems.

\section*{Acknowledgments}

Ang\'{e}lica S. Mata acknowledges FAPEMIG (Grant No. APQ-02482-18) and CNPq (Grant No.  423185/2018-7). Guilherme F. de Arruda, Yamir Moreno, and Lucas Jeub acknowledge support from Intesa Sanpaolo Innovation Center. Yamir Moreno acknowledges partial support from the Government of Arag\'on and FEDER funds, Spain through grant ER36-20R to FENOL, and by MINECO and FEDER funds (grant FIS2017-87519-P). The funders had no role in study design, data collection and analysis, decision to publish, or preparation of the manuscript. Francisco Rodrigues acknowledges CNPq (grant 309266/2019-0) and FAPESP (grant 19/23293-0) for the financial support given for this research. Research carried out using the computational resources of the Center for Mathematical Sciences Applied to Industry (CeMEAI) funded by FAPESP (grant 2013/07375-0).

\appendix
\section{Monte Carlo simulations}
\label{app:MC}

In this section, we will focus on the computational viewpoint to model the rumor process described in the main text. Denoting by $N_i$ the number of spreaders, $N_k$ the number of edges emanating from spreaders, and $N_r$ the number of stiflers. We implemented our model as follows. At each step, with probability $\delta N_r/[\delta N_r+ \lambda N_k + \alpha N_k ]$ one stifler, chosen at random, forgets the rumor and becomes an ignorant again. With probability $(\lambda + \alpha) N_k/[\delta N_r+ \lambda N_k + \alpha N_k ]$ contact is made. Algorithmically, this contact is implemented in two steps: (i) A spreader vertex $i$ is selected with probability proportional to its degree, next (ii) a nearest neighbor of $i$, here denoted as $j$, is selected uniformly. If $j$ is an ignorant, with probability $\lambda/[\lambda + \alpha]$, $j$ learns the rumor and becomes a spreader. If $j$ is another spreader or a stifler, with probability $\alpha/[\lambda + \alpha]$ it becomes a stifler. If none of these conditions are satisfied, nothing happens.  Next, time is incremented by $dt = 1/[N_r+(\lambda N_k)+(\alpha N_k)]$. 

With this scheme, we have two different algorithms, one for $\delta > 0$, where we have a single absorbing state (rumor-free), and another for $\delta = 0$, where we have a strong dependency on the initial condition and many absorbing states. In the former case, the simulations were performed using the quasi-stationary method ~\cite{DeOliveira05}, which is one of the most robust approaches to overcome the stationary simulations' difficulties of finite systems with absorbing states. In this method, every time the process tries to visit an absorbing state, this state is substituted by an active configuration previously visited during the simulation. These active configurations are stored in a list, which is constantly updated and works as a new initial condition. This approach is completely equivalent to the standard quasi-stationary method where averages are performed only over samples that did not visit the absorbing state~\cite{Marrobook}. In our experiments, we let the simulations run during a relaxation time $t_r = 10^7$ time steps and computed the averages over a time $t_{av} =  10^7$. The threshold can be estimated using the modified susceptibility~\cite{Ferreira2012},
\begin{equation}
 \chi = \frac{ \E{ (n^*)^2 } - \E{ n^* }^2 }{\E{ {n^*} }} = N \left( \frac{ \E{ (\rho^{QS})^2 } - \E{ \rho^{QS} }^2 }{\E{ \rho^{QS} }} \right),
\end{equation}
where $n^*$ is the number of spreaders, and $\rho^{QS}$ is the quasi-stationary distribution. As argued in~\cite{Ferreira2012, Mata2015, Arruda2018}, the susceptibility presents a peak at the phase transition in finite systems. The parameters corresponding to the maximum value of the susceptibility will coincide with the critical threshold for sufficiently large systems. On the other hand, for the $\delta = 0$ case, the algorithm consists of running the simulation, beginning with a single randomly placed seed and calculating the final fraction of stiflers. To have a better estimation, we run in the range of $500$ up to $5 \times 10^5$ independent simulations, guaranteeing that $\chi$ does not vary more than $10^{-3}$ if compared with the result before the last batch of $500$ simulations. We highlight that, although in this case, we are not interested in the susceptibility, as it depends on the second moment of the distribution of stiflers, this will guarantee a reasonable sampling for the first moment and its peak was also a reasonable indicator of the transition as well. We also remark that, in our experiments, the susceptibility peak coincides with the lifespan peak calculated using the method proposed in Ref.~\cite{Boguna2013}.

\section{Complementary simulation results}

\subsection{Maki--Thompson model on random regular networks}
\label{app:rrn}

\begin{figure}[t!]
    \includegraphics[width=\linewidth]{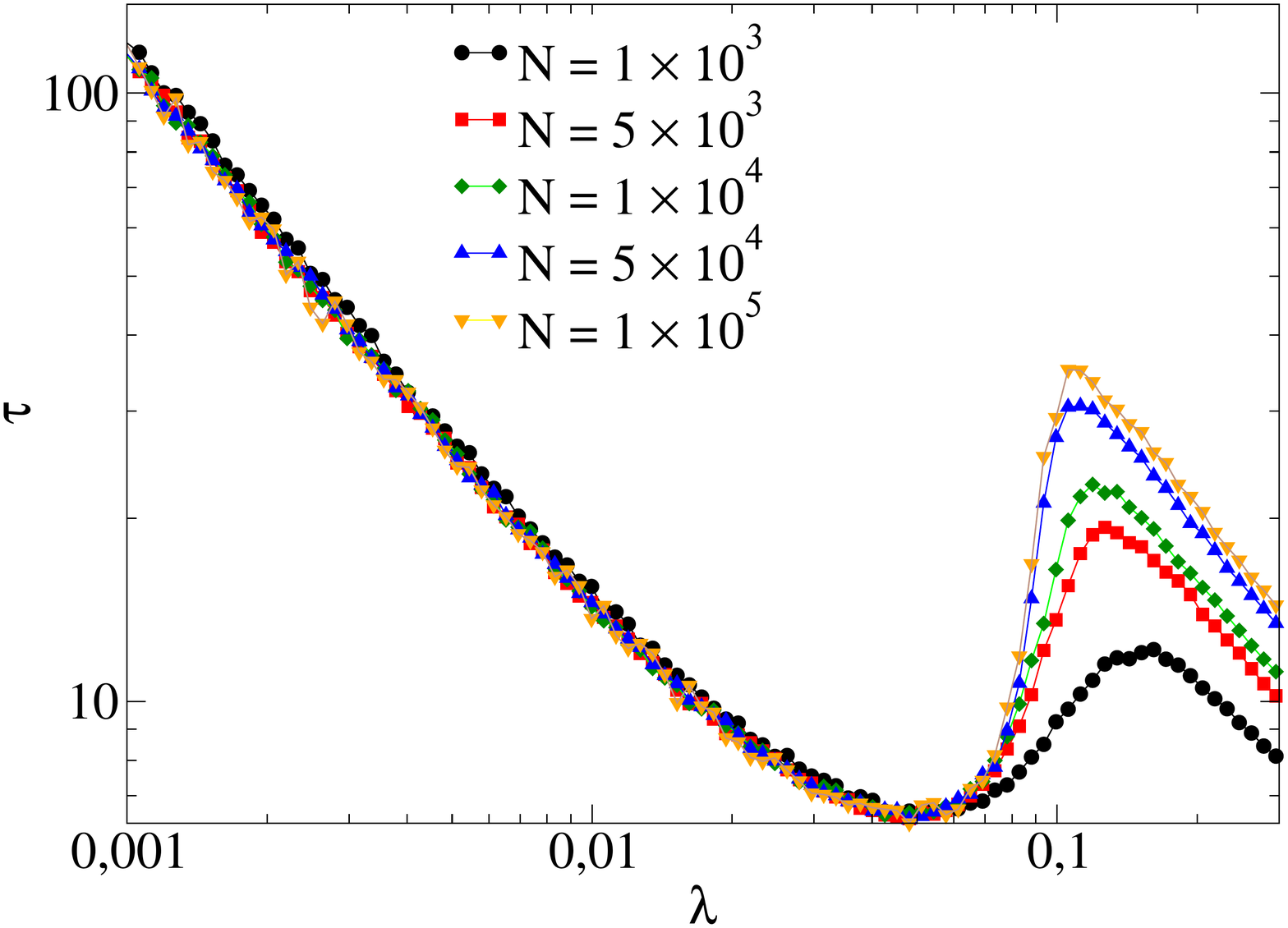}
    \caption{Time to reach the absorbing state for the standard MT model, $\alpha = 1$ and different sizes on a random regular networks with $\langle k \rangle = 10$. For the phase diagram, please see Fig.~\ref{fig:MK_critical} in the main text.}
    \label{fig:MK_T}
\end{figure}

Complementary to the phase diagram reported in the main text, in Fig.~\ref{fig:MK_T}, we show their corresponding lifespan. We note that the subcritical regime is still present and follows a similar pattern, as observed in Fig.~\ref{fig:Tabs_loglog}. Additionally, note that we can also observe a peak in the lifespan near the second-order phase transition.

\subsection{Power-law networks and $\delta = 1$}
\label{app:PL}

\begin{figure}[t!]
    \includegraphics[width=\linewidth]{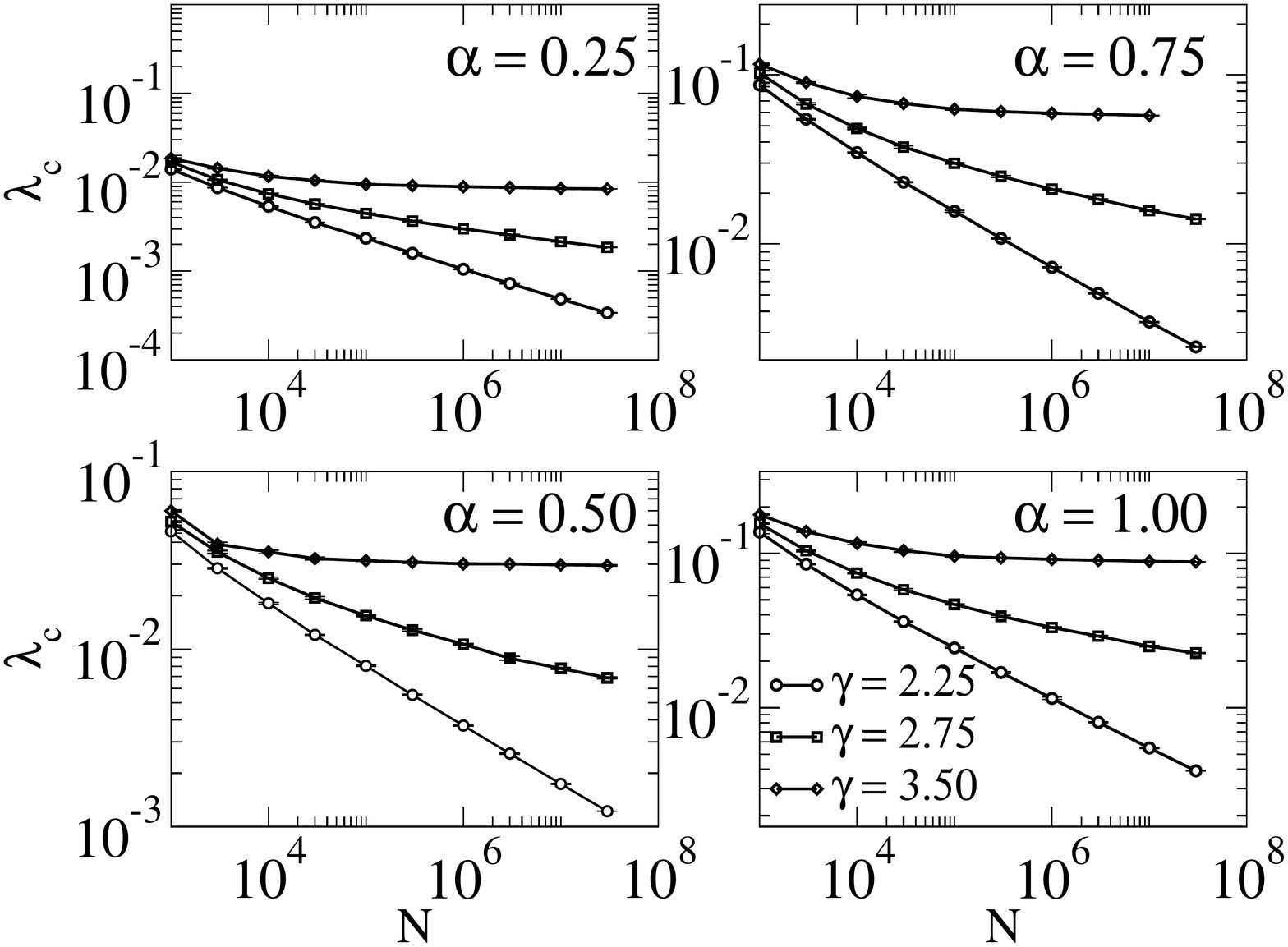}
    \caption{Critical point estimations of uncorrelated power-law networks as a function of $N$ and for different values of $\gamma$ and $\alpha$.}
    \label{fig:ucm_rumor_lbc}
\end{figure}

\begin{figure}[ht]
    \includegraphics[width=\linewidth]{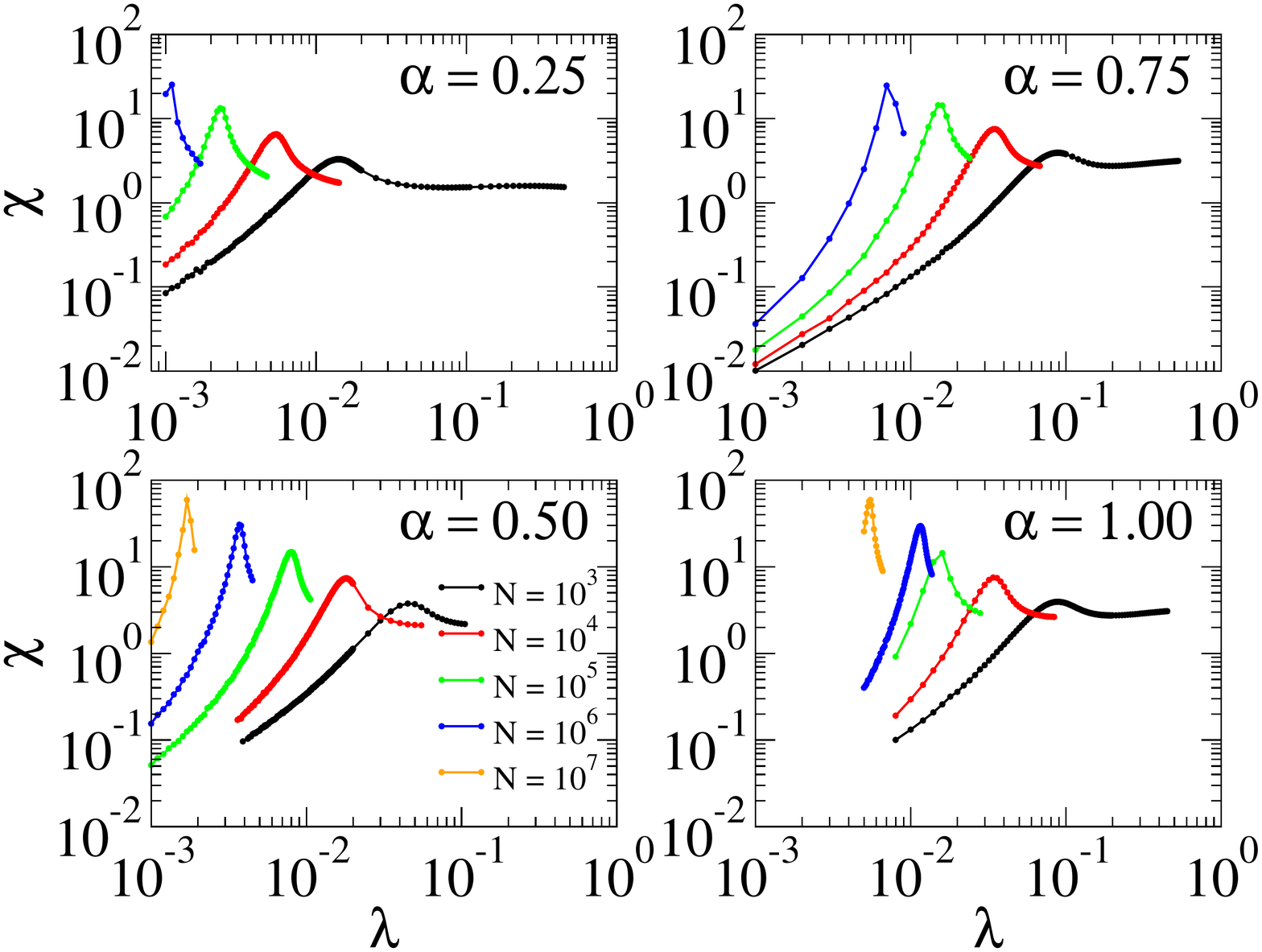}
    \caption{Susceptibility curves for different sizes and as a function of $\lambda$ for $\alpha = 0.25, 0.5, 0.75$ and $1.0$, considering power-law networks, $P(k) \sim k^{-\gamma}$, with $\gamma \approx 2.25$.}
    \label{fig:ucm_g225rumor}
\end{figure}

\begin{figure}[ht]
    \includegraphics[width=\linewidth]{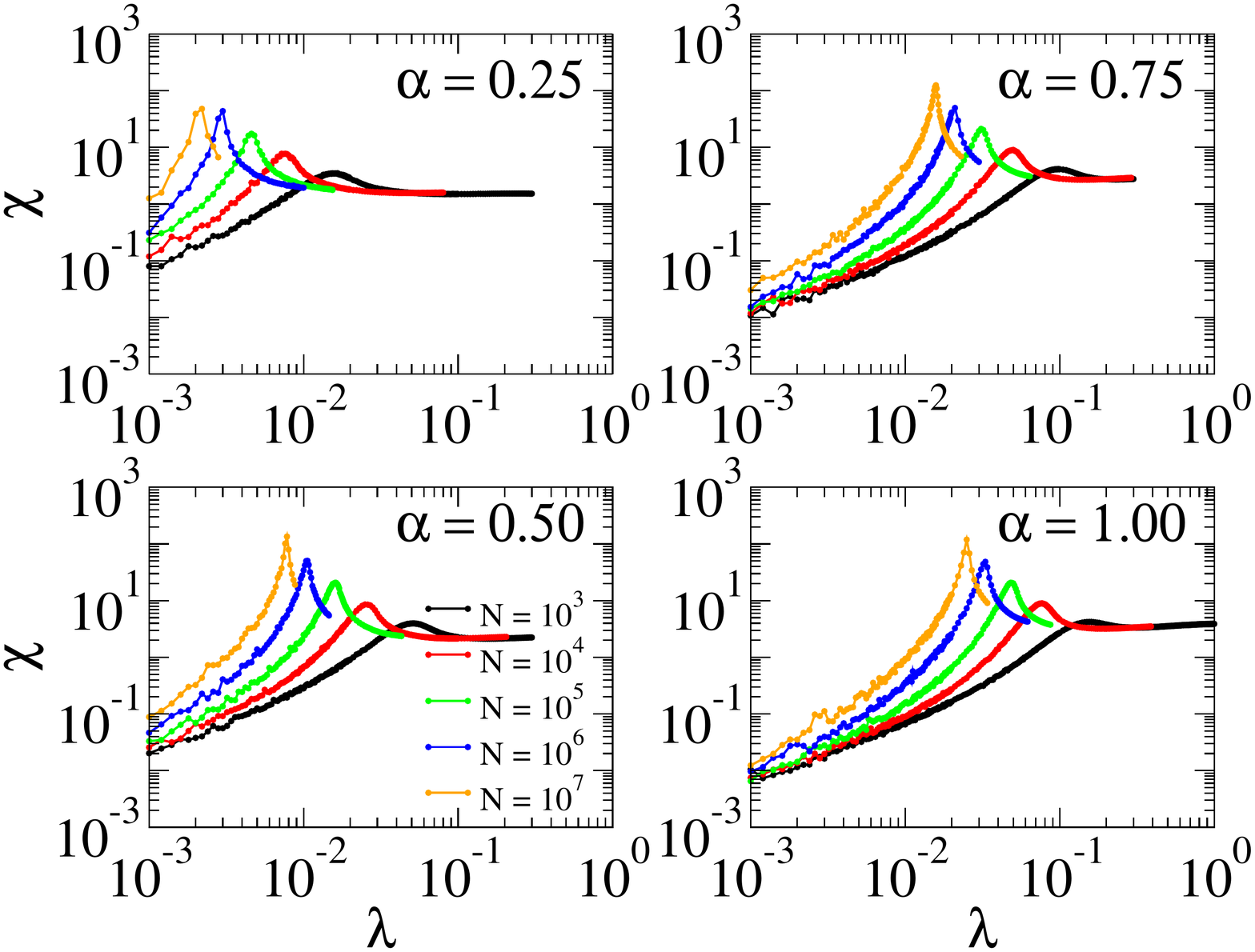}
    \caption{Susceptibility curves for different sizes and as a function of $\lambda$ for $\alpha = 0.25, 0.5, 0.75$ and $1.0$, considering power-law networks, $P(k) \sim k^{-\gamma}$, with $\gamma \approx 2.75$.}
    \label{fig:ucm_g275rumor}
\end{figure}

\begin{figure}[ht]
    \includegraphics[width=\linewidth]{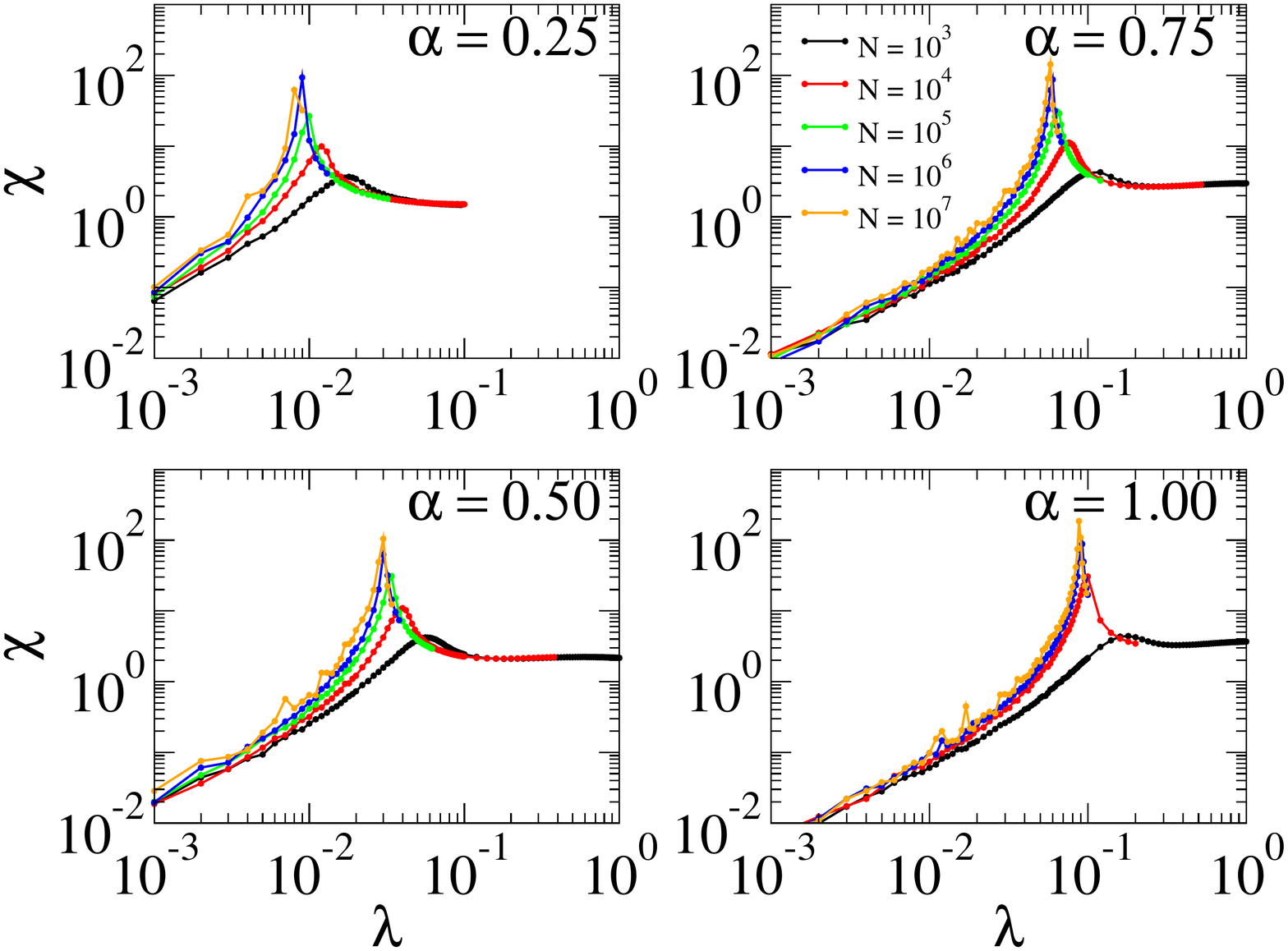}
    \caption{Susceptibility curves for different sizes and as a function of $\lambda$ for $\alpha = 0.25, 0.5, 0.75$ and $1.0$, considering power-law networks, $P(k) \sim k^{-\gamma}$, with $\gamma \approx 3.5$.}
    \label{fig:ucm_g350rumor}
\end{figure}

\begin{figure}[ht]
    \includegraphics[width=\linewidth]{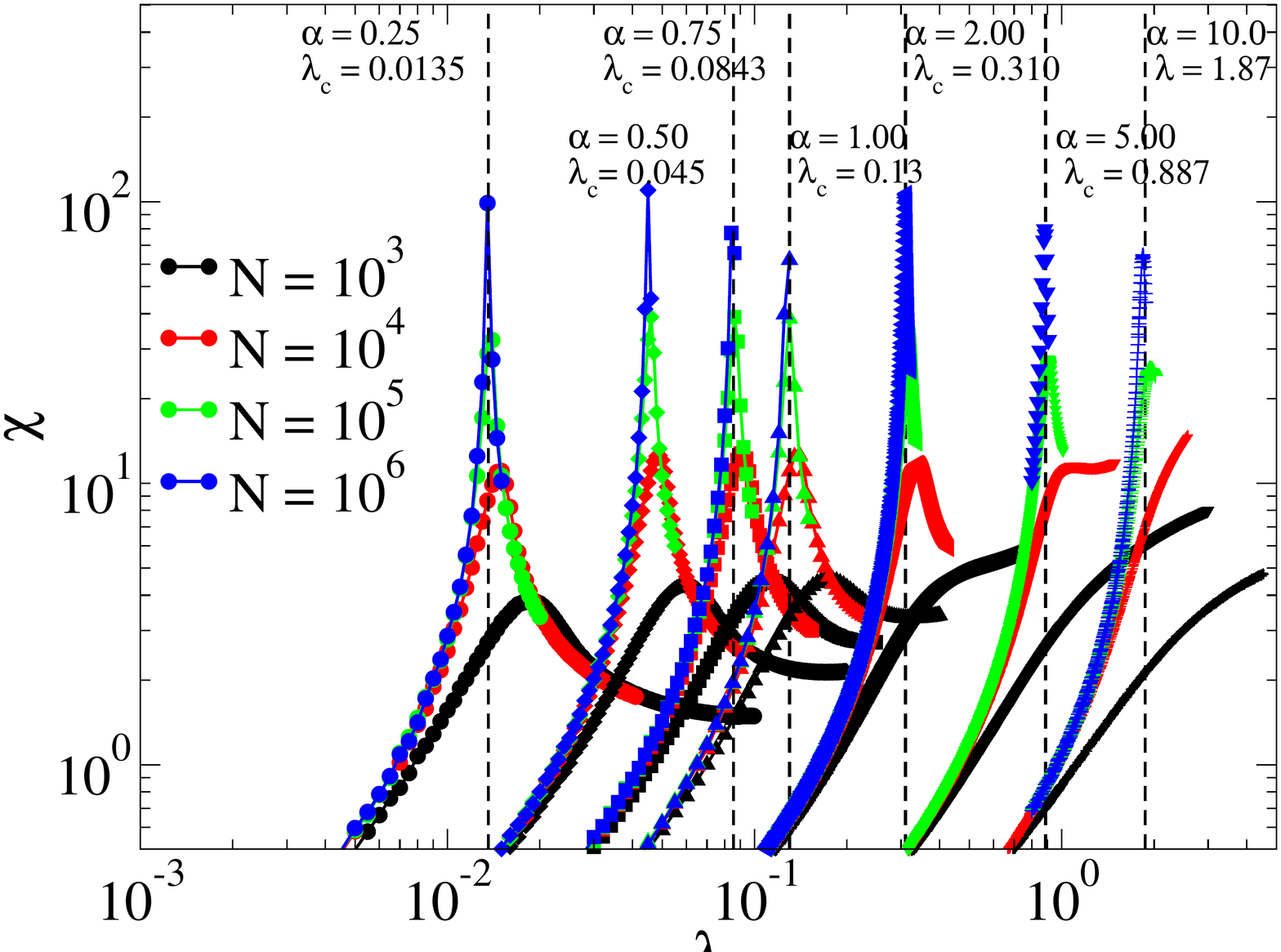}
    \caption{Susceptibility curves as a function of $\lambda$ for random regular networks with $\langle k \rangle = 10$ and different values of $\alpha$ and sizes.}
    \label{fig:rrn_10}
\end{figure}

Complementary to the homogeneous cases presented in the main text, in Fig.~\ref{fig:ucm_rumor_lbc} we show the finite-size scaling for power-law networks, $P(k) \sim k^{-\gamma}$, for different values of $\alpha$. The values of $\lambda_c$ are obtained using the Figs.~\ref{fig:ucm_g225rumor},\ref{fig:ucm_g275rumor} and \ref{fig:ucm_g350rumor} where the susceptibility in function of $\lambda$ was calculated for different uncorrelated power-law network sizes. First, we observe that, as $\alpha$ increases, the critical point also increases. Additionally, for $\gamma = 2.25$, our experiments suggest that the critical point vanishes in the thermodynamic limit. On the other hand, for $\gamma = 3.5$, we observe that the critical point converges to a non-zero value. For $\gamma = 2.75$, from our experiments, a vanishing behavior is reasonable. As mentioned in the main text, this behavior is at odds with the behavior of the SIS, in which the power-law networks have a vanishing critical point for any value of $\gamma$~\cite{chatterjee2009, montford2013, Arruda2018}. On the other hand, it follows a similar pattern as for the SIRS model~\cite{Ferreira2016}, contact process \cite{cp}, the generalized SIS model with weighted infection rates~\cite{KJI} and also modified versions of the SIS model~\cite{COTA2018}.

\section{Asymptotic analysis}
\label{app:aa}

In this section, we present the details of the asymptotic analysis. The first-order approximation assumes that the states of the nodes are independent, hence, the set of equations that describe its temporal evolution are given as
\begin{align} \label{eq:qmf_app}
        \frac{d x_i}{dt} =& \delta z_i - \lambda \sum_{k=1}^N \A_{ki} x_i y_k \\
        \frac{d y_i}{dt} =& \lambda \sum_{k=1}^N \A_{ki} x_i y_k - \alpha \sum_{k=1}^N \A_{ki} y_i \left(y_k + z_k \right) \\
        \frac{d z_i}{dt} =& -\delta z_i + \alpha \sum_{k=1}^N \A_{ki} y_i \left(y_k + z_k \right). 
\end{align}
Following an asymptotic analysis, we consider that $y_i = y_i^{(1)} \epsilon^c + O(\epsilon^{2c})$, $z_i = z_i^{(1)} \epsilon^k + O(\epsilon^{2k})$ and $x_i \in O(1)$, where $\epsilon \ll 1$. Moreover, in order to balance the equations, we have to consider scaling the parameters as $\lambda = \tilde{\lambda} \epsilon^m$ and $\alpha = \tilde{\alpha} \epsilon^n$. Without loss of generality, we also assume $\delta = 1$. In this formulation, the scaling of the parameters will be related with the scaling of the nodal probabilities. From Eq.~\eqref{eq:qmf_app} in the steady-state, where the derivatives tend to zero, and neglecting the higher-order terms, we have
\begin{align} \label{eq:aa_app}
        \delta z_i^{(1)} \epsilon^k &= \lambda \epsilon^m \sum_{k=1}^N \A_{ki} x_i^{(1)} y_k^{(1)} \epsilon^c\\
        \delta z_i^{(1)} \epsilon^k &= \alpha \epsilon^n \sum_{k=1}^N \A_{ki} y_i^{(1)} \epsilon^c \left(y_k^{(1)} \epsilon^c + z_k^{(1)} \epsilon^k \right).
\end{align}
Note that, to respect the order of the terms in both sides of Eq.~\ref{eq:aa_app} we also have to respect $k = n + c + \min(c, k)$ and $k = m + c$. This conditions imply that we have three different regimes:  (i) $0 < c < k$, where $c = m -n$ and $k = 2m -n$, discussed in Sec.~\ref{app:regime_0ck} (ii) $c = k$, where $m = 0$ and $n = -k$, discussed in Sec.~\ref{app:regime_ck}, and (iii) $0 < k < c$, where $c = -n$ and $k = m - n$, discussed in Sec.~\ref{app:regime_0kc}. 

As previously mentioned in the main text, this analysis emphasizes that we cannot obtain the critical point at a first-order approximation. However, we remark that the phase transition was observed in the Monte Carlo simulations. For this reason, we call this an elusive transition. Thus, this is a limitation of our mean-field approximations, suggesting that different approximations might be necessary to properly estimate the critical point.

\subsection{Regime $0 < c < k$}
\label{app:regime_0ck}

As previously mentioned, the parameters and the order of the nodal probabilities are not independent. Considering only the dominating terms of Eq.~\ref{eq:aa_app} in the regime $0 < c < k$, we have that
\begin{align}
    z_i^{(1)} &= \tilde{\lambda} \sum_j \A_{ij} y_j^{(1)} \label{eq:0ck_0} \\
    z_i^{(1)} &= \tilde{\alpha} \sum_j \A_{ij} y_i^{(1)} y_j^{(1)}. \label{eq:0ck_1}
\end{align}

Defining $\bm y = \left[ y_1^{(1)}, y_2^{(1)}, ..., y_N^{(1)} \right]^T$ and plugging Eq.~\ref{eq:0ck_0} into Eq.~\ref{eq:0ck_1} we have
\begin{equation}
 \tilde{\lambda} \A \bm y = \tilde{\alpha} \bm y \circ (\A \bm y),
\end{equation}
where $\circ$ denotes the Hadamard (i.e., element-wise) product. Consequently, this relation yields
\begin{equation}
    y_i^{(1)} = \frac{\tilde{\lambda}}{\tilde{\alpha}}.
\end{equation}
Substituting the scaling relationships, this is equivalent to
\begin{equation}
y_i = \frac{\lambda}{\alpha} \epsilon^{n-m+c} + O(\epsilon^{2c})=\frac{\lambda}{\alpha} + O(\epsilon^{2c}),
\end{equation}
which shows that in this regime, at leading order, the probability of each node being a spreader depends only on $\lambda$ and $\alpha$. Moreover, for positive values of $\lambda$ and $\alpha$, $y_i$ will be positive and does not depend on the network.

\subsection{Regime $c = k$}
\label{app:regime_ck}

Next, considering only the dominating terms of Eq.~\ref{eq:aa_app} in the regime $c = k$, we have
\begin{align}
    z_i^{(1)} &= \tilde{\lambda} \sum_j \A_{ij} y_j^{(1)}, \label{eq:ck_0}  \\
    z_i^{(1)} &= \tilde{\alpha} \sum_j \A_{ij} y_i^{(1)} \left( y_j^{(1)} + z_j^{(1)} \right). \label{eq:ck_1} 
\end{align}

Similar to the previous section, plugging Eq.~\ref{eq:ck_0} into Eq.~\ref{eq:ck_1}, we obtain
\begin{equation}
\label{eq:ck}
 \tilde{\lambda} \A \bm y = \tilde{\alpha} \bm y \circ (\A \bm y + \tilde{\lambda}\A^2 \bm y).
\end{equation}
We do not have a closed-form solution for Eq.~\ref{eq:ck} for general networks. However, if we assume that $\A$ represents a regular network and $\bm y$ is a constant vector, we have $\A \bm y = \Lambda_{\max} \bm y$, where $\Lambda_{\max}$ is the leading eigenvalue of $\A$. Hence,
\begin{equation*}
\tilde{\lambda}\Lambda_{\max} y_i^{(1)} = \tilde{\alpha} y_i^{(1)} (\Lambda_{\max} + \tilde{\lambda} \Lambda^2_{\max}) y_i^{(1)},
\end{equation*} 
which yields
\begin{equation}
 y_i^{(1)} = \frac{\tilde{\lambda}}{\tilde{\alpha} \left( \tilde{\lambda}  \Lambda_{\max} + 1 \right)}
\end{equation}
and
\begin{equation}
 z_i^{(1)} = \frac{\tilde{\lambda}^2 \Lambda_{\max} }{\tilde{\alpha} \left( \tilde{\lambda}  \Lambda_{\max} + 1 \right)}.
\end{equation}
Substituting the scaling relationships as before, we obtain
\begin{align}
y_i &= \frac{\lambda}{\alpha(\lambda \Lambda_{\max} + 1)} + O(\epsilon^{2c}),\\
z_i &= \frac{\lambda^2 \Lambda_{\max}}{\alpha (\lambda \Lambda_{\max} + 1)} + O(\epsilon^{2c}).
\end{align}
As in the previous regime, $y_i > 0$ for positive values of $\lambda$ and $\alpha$. Note that in the regime $c = k$ there is a dependency on the network structure, here codified in the leading eigenvalue. However, such a dependency does not define a critical point.

\subsection{Regime $0 < k < c$}
\label{app:regime_0kc}

Finally, in the last case, we consider the dominating terms of Eq.~\ref{eq:aa_app} for the regime $0 < k < c$. Thus, we have
\begin{align}
    z_i^{(1)} &= \tilde{\lambda} \sum_j \A_{ij} y_j^{(1)}, \\
    z_i^{(1)} &= \tilde{\alpha} \sum_j \A_{ij} y_i^{(1)} z_j^{(1)}.
\end{align}
Following the same approach as before, we have
\begin{equation}
 \tilde{\lambda} \A \bm y = \tilde{\alpha} \bm y \circ (\A^2 \bm y).
\end{equation}

Next, assuming that $\A$ represents a regular network and using the same argument as in previous section, we have
\begin{equation}
 y_i = \frac{1}{\alpha \Lambda_{\max}} + O(\epsilon^{2c}).
\end{equation}
Again we observe that $y_i > 0$. However, in this regime this condition depends only on $\alpha$ and the network structure through $\Lambda_{\max}$.

\end{document}